\documentclass{article}
\usepackage[utf8]{inputenc}
\usepackage{graphicx}
\usepackage{caption}
\usepackage{subfig}
\usepackage{amsmath}
\usepackage[version=4]{mhchem}
\usepackage{siunitx}
\usepackage{longtable,tabularx}
\usepackage{booktabs}
\usepackage{wrapfig}
\usepackage{fullpage}
\title{Wind Tunnel Tests of a Pitot-Static Tube Array to Estimate Wind Velocity}

\author{Matthew T. Simmons\footnote{Graduate Researcher, Department of
    Mechanical Engineering, University of South Alabama.}, Carlos
  J. Montalvo\footnote{Assistant Professor, Department of Mechanical
    Engineering, University of South Alabama.}, and Sytske
  K. Kimball\footnote{Chair, Department of Earth Sciences, University
    of South Alabama} \\
Department of Mechanical Engineering \\ Facility for Aerial Systems and
Technology\\University of South Alabama, Mobile, AL, 36608 USA}

\begin{document}
\maketitle

\begin{abstract}
This paper examines the use of a pitot-static tube array (PSA) to estimate horizontal wind velocity in all directions. This method uses the readings from the PSA as inputs. No flight tests have been done, but this method could potentially be used to estimate wind velocity from small UAVs. To do this, UAV ground speed and attitude must be available as inputs. Wind tunnel data of the PSA was used to create an algorithm that estimates wind velocity even when the wind is not parallel with any of the pitot-static tubes. This paper discuses the wind tunnel data, the algorithm, and the application for small UAVs.
\end{abstract}

\section{Introduction}
When measuring airspeed, pitot-static tubes are typically used to take measurements along the longitudinal axis of the aircraft. Five hole and seven hole probes can take airspeed measurements up to 75 degrees from the aircraft's longitudinal axis allowing angle of attack and side slip angle to be determined.\cite{pisasale2003examining}\cite{pisasale2002novel} Other devices such as optical flow and alpha-beta vanes can be used to make these measurements.\cite{elston2015overview}\cite{gracey1958summary} However, none of these devices can measure airspeed up to 180 degrees from the aircraft's longitudinal axis which is required to measure wind velocity in all directions. Sensors such as twelve hole probes, thermal anemometers, and sonic anemometers can measure airspeed in all directions (nearly all directions in the case of the twelve hole probe).\cite{ramakrishnan2007development}\cite{matayoshi2005development} For the application of a cheap and small UAV, these devices are too heavy and costly creating a need for a low cost and light weight solution. A 3D printed version of a five hole probe has been made, but it is only able to measure wind velocity up to 25 degrees from the longitudinal axis.\cite{azartash2017evaluation} As 3D printing technology improves, it may be feasible in the future to print cheap 12 hole probes. Methods to estimate horizontal wind velocity using a single pitot-static probe along its longitudinal axis while flying in a curved path have been developed for meteorological UAVs.\cite{johansen2015estimation}\cite{mayer2012no}\cite{langelaan2011wind} The method presented in this paper uses a pitot-static tube array (PSA) to provide a low cost and light weight solution to estimate horizontal wind velocity. This method would not require a specific flight path if used on a UAV. The concept and structure of the PSA originated in work published on using a fleet of UAVs for atmospheric characterization. \cite{lisa} This work used a pitot-static tube array consisting of four pitot-static tubes to measure scalar wind speed in each cardinal direction. This work paved the way for the PSA presented in this paper; they determined the best installation position to avoid down wash effects, calibration procedures, and performed UAV flight tests. Sensirion, a sensor manufacture headquartered in Switzerland, used a similar approach to measure horizontal wind velocity.\cite{sensirion} They 3D printed a relatively small and light weight disk with static ports and stagnation ports on the side that would take the place of the pitot-static tubes in the PSA. Their static ports and stagnation ports are 180 degrees apart as opposed to 90 degrees apart with a typical pitot-static tube. They do not discuss in detail the method or results. Table \ref{table:payload_cost} shows the weights and cost ranges of a few onmidirectional wind velocity devices compared to the PSA. The cost range of each device covers a range of accuracies and manufacturers. The weights and costs of the required microcontroller, GPS module, IMU module, and circuit board are not included in Table \ref{table:payload_cost}. These components are required no matter which device is used to determine wind velocity.
\begin{table}[!ht]
\centering
\begin{tabular}{|c|c|c|c|c|}
	\hline
	\textbf{Device} & \textbf{Cost Range} & \textbf{Approximate Weight} & \textbf{Voltage Required} \\
    \hline
	Twelve Hole Probe & \$5,000-\$7,000 & 100 g & 5 VDC \\
    \hline
	Thermal Anemometer & \$100-\$2000 & 400 g & 6 VDC \\
	\hline
	Sonic Anemometer & \$300-\$3000 & 400 g & 10 VDC \\
	\hline
	Pitot-Static Tube Array & \$150-\$250 & 150 g & 5 VDC \\
	\hline
\end{tabular}
\caption{Comparison of wind velocity devices.}
\label{table:payload_cost}
\end{table}      
\newpage
\section{Pitot-Static Tube Array}
\begin{figure}[!ht]
  \begin{center}
  \includegraphics[width=0.5\textwidth]{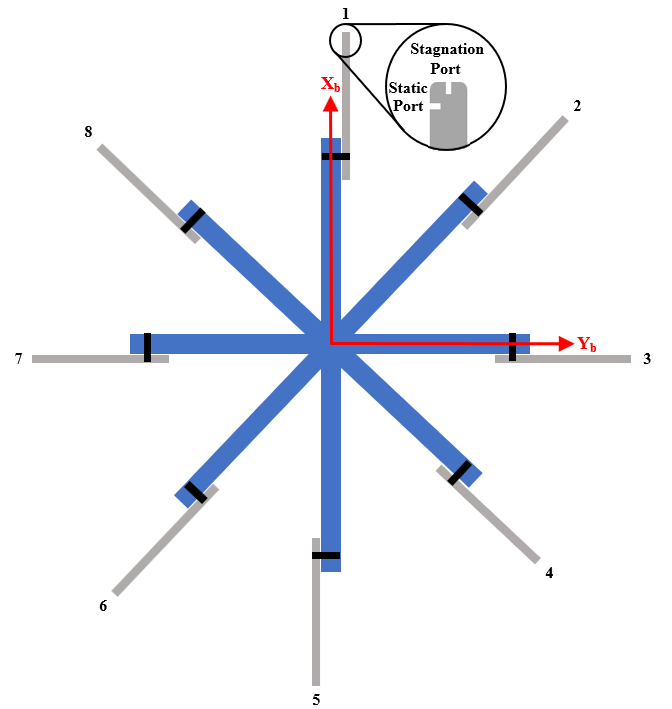}
  \caption{PSA orientation diagram.}
  \label{fig:FP8_orientation}
  \end{center}
\end{figure}

The PSA sensor is constructed using eight individual pitot-static tubes horizontally orientated in 45 degree intervals about a common vertical axis. A 3D printed eight spoke frame is used to fasten the pitot-static tubes in place. The pitot-static tubes are connected to individual membrane style differential pressure sensors  which are wired into an Arduino DUE controller. The sensors are model MPXV7002DP made by NPX. The differential pressure sensor data for each pitot-static tube is written to a text file on a micro SD memory card by the Arduino controller at a sample rate of 5 $Hz$. The body reference frame of the PSA is shown in red in Figure \ref{fig:FP8_orientation}.

\section{Wind Tunnel Tests}
\begin{figure}[!ht]
  \begin{center}
  \includegraphics[width=0.5\textwidth]{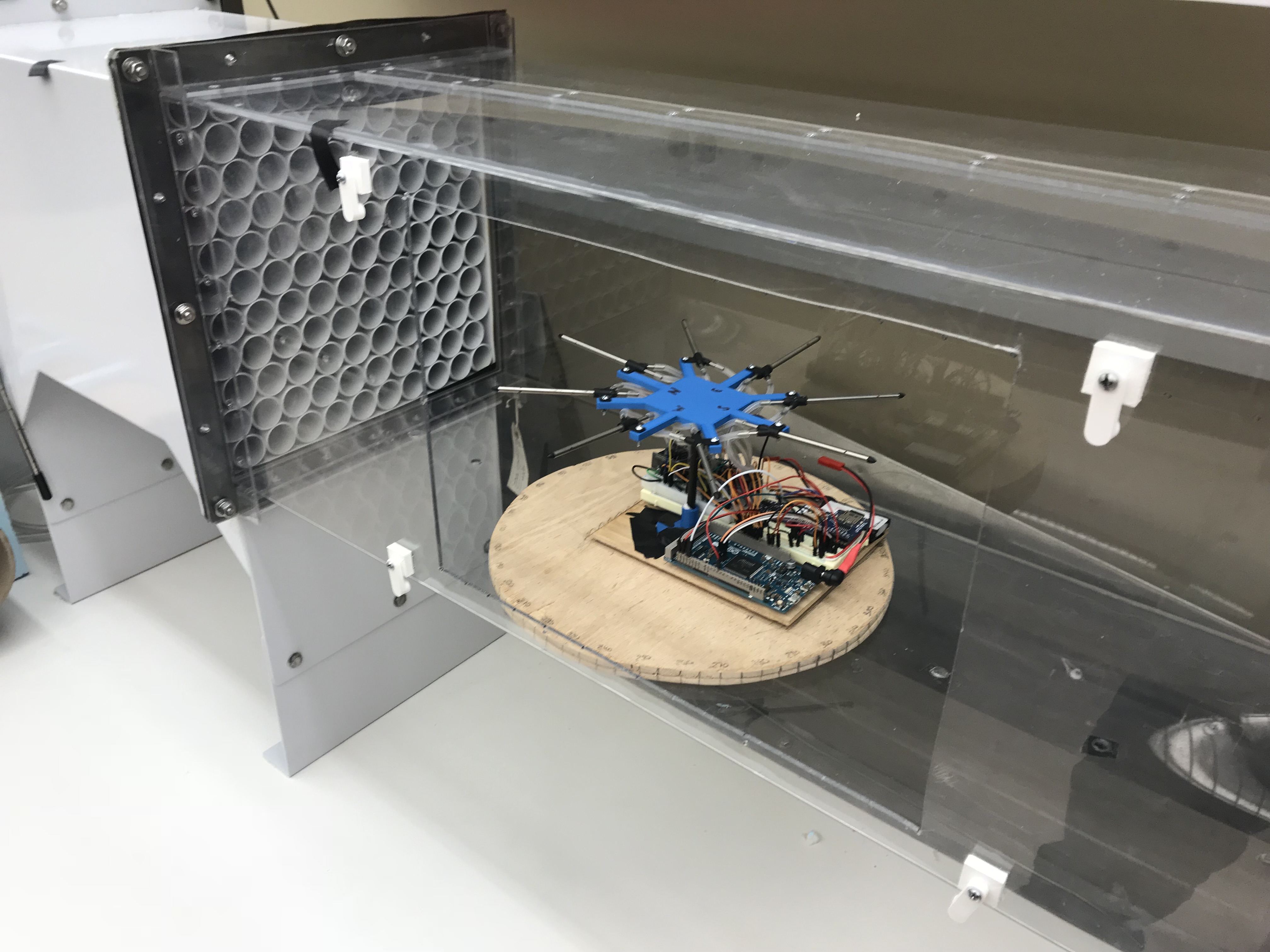}
  \caption{PSA in the wind tunnel.}
  \label{fig:FP8_inTunnel}
  \end{center}
\end{figure}

\begin{figure}[!ht]
  \centering
  \subfloat[Measured wind speed as the pitot-static tube's yaw angle is rotated.]{\includegraphics[width=0.4\textwidth]{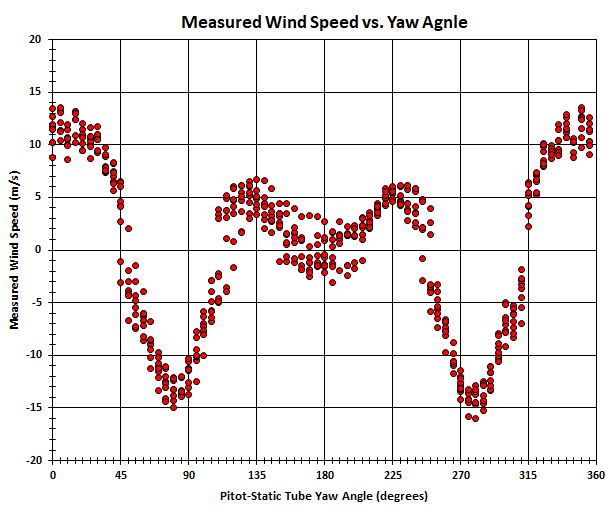}\label{fig:dataPlot}}
  \subfloat[Normalized data and 4th order Fourier series fit.]{\includegraphics[width=0.4\textwidth]{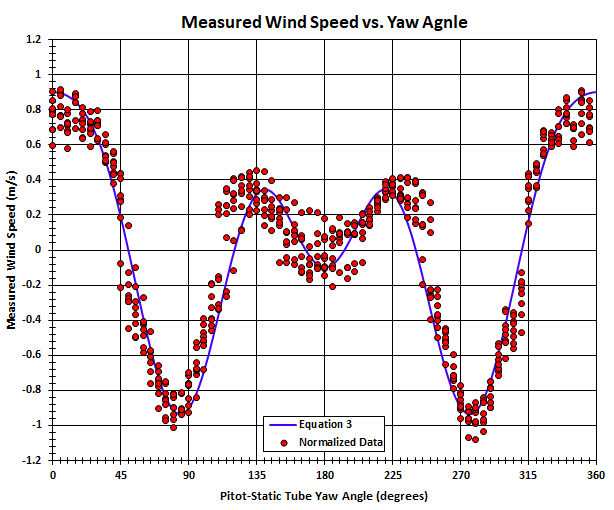}\label{fig:normPlot}}
  \caption{Results from wind tunnel testing.}
  \label{fig:tunnelData}
\end{figure}

The PSA sensor was placed in a wind tunnel at the University of South Alabama, and data was recorded at a constant wind speed while the PSA yaw angle was rotated in five degree intervals from 0 to 360 degrees. Each yaw angle was held for 60 seconds, and averages from each 60 second interval were taken. In total from eight pitot static tubes, 584 averages were taken. In Figure \ref{fig:FP8_inTunnel}, the wooden round disk is the platform used to rotate the PSA. A shaft fastened to the bottom of the disk and through a small hole in the bottom of the wind tunnel is used to manually rotate the assembly. Five degree tick marks are labeled on the side of the disk. These are used with a stationary mark on the wind tunnel to align the assembly throughout the rotation intervals. 
The first 60 seconds of recorded data was taken with the wind tunnel off. This data is used to calibrate each sensor so that the ambient pressure reading is the same for all. For this experiment, all raw bit data is scaled to have a value of 520 at ambient pressure. After the first 60 seconds, the wind tunnel generated 14.8 $m/s$ wind as measured by a factory calibrated hand held wind gauge. The averages taken from each 60 second interval exclude the data from transitions while the PSA was rotated. Averages were taken from approximately 45 second durations within each 60 second interval. The raw bit value for each average is converted to pressure using Equation \ref{eq:pressure}.
\begin{equation}
\label{eq:pressure}
P_{s}=(\frac{3.3}{1023})(r_{s}-r_{a})/P_{a} 
\end{equation}
\noindent
Where $P_{s}$ is pressure, $r_{s}$ is the raw bit value, $r_{a}$ is the raw bit value at ambient pressure, $P_{a}$ is ambient pressure, and $\frac{3.3}{1023}$ is a factor specific to the Arduino DUE controller. Then, wind speed is calculated using Equation \ref{eq:windspeed}.\cite{windspeed} 
\begin{equation}
\label{eq:windspeed}
U=\sqrt{5(\gamma{R}T)((P_{s}+1)^{\frac{2}{7}}-1)}
\end{equation}
\noindent
Where $U$ is wind speed, $\gamma$ is the adiabatic index of air, $R$ is the molar gas constant, $T$ is ambient temperature, and $P_{s}$ is the pressure from Equation \ref{eq:pressure}. 
\newline
\newline
\indent
The data taken shows a sinusoidal relationship between the measured wind speed of a pitot-static tube and its yaw angle with respect to the wind. Figure \ref{fig:dataPlot} shows the 584 data points taken. In Figure \ref{fig:dataPlot}, the x-axis is the yaw angle of each individual pitot-static tube at the time the data point was taken and not the yaw angle of the PSA as a whole. When the pitot-static tube is rotated 90 degrees the static port becomes the stagnation port and the stagnation port becomes the static port. Therefore, at a yaw angle of 90 degrees, a pitot-static tube will measure the full magnitude of the wind speed, but negative because the high side and low side of the differential pressure sensor remain the same. This occurs at yaw angles of 90 degrees and 270 degrees. In Figure \ref{fig:dataPlot}, these events are shown at approximately 85 degrees and 275 degrees. This is most likely due to the interval markings on the rotation disk in the wind tunnel not being perfectly aligned. The measured wind speeds at 0 degrees and 360 degrees are approximately 1 $m/s$ less than the magnitude of the measured wind speeds at 90 degrees and 270 degrees. This is most likely due to the pitch angle of the PSA sensor not being exactly 0 degrees. Although these discrepancies exist, the data can still be used as proof of concept for the methods presented in this paper. 
\newline
\newline
\indent
To make the plot applicable for all wind speeds, the measured values were normalized to one by dividing by 14.8 $m/s$, the actual wind speed in the tunnel. From the normalized data, a 4th order Fourier series fit can be approximated to describe the relationship. Using a least squares fit, the first five coefficients in the series were approximated. The series fit is given by Equation \ref{eq:seriesFit}.
\begin{equation}
\label{eq:seriesFit}
\hat{u}(\psi) = \frac{a_{o}}{2}+\sum_{n=1}^{4}a_{n}\cos\left(\frac{n\pi\psi}{L}\right) 
\end{equation}
\noindent
Where $\hat{u}(\psi)$ is the normalized measured wind speed, $\psi$ is the yaw angle of the pitot-static tube with respect to the wind, $L$ is the period, and $a_{o}$, $a_{1}$, $a_{2}$, $a_{3}$, $a_{4}$ are the term coefficients. Figure \ref{fig:normPlot} shows the normalized data compared to Equation \ref{eq:seriesFit}. The coefficient of determination is $R^2 = 0.78$.

\section{Estimating Wind Speed and Wind Direction}
\begin{figure}[!ht]
  \centering
  \subfloat[Wind direction relative to the PSA's x-axis.]{\includegraphics[width=0.4\textwidth]{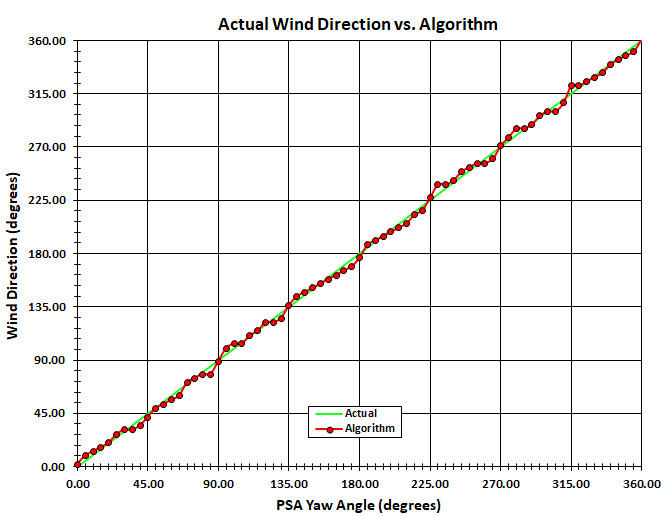}\label{fig:direction8}}
  \subfloat[Wind speed.]{\includegraphics[width=0.4\textwidth]{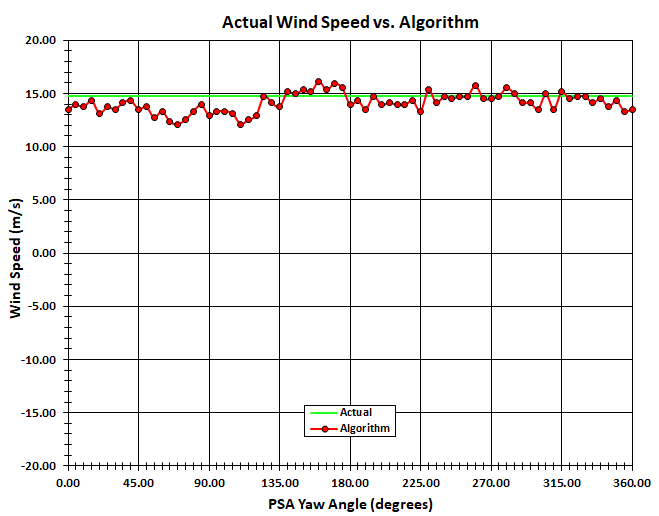}\label{fig:speed8}}
  \caption{Algorithm's outputs compared to actual conditions during the wind tunnel test.}
  \label{fig:algorithm8}
\end{figure}

A grid search method algorithm was written to determine the most likely wind speed and wind direction given the inputs of the eight pitot-static tubes at any given time. For each set of inputs at any given time, the algorithm uses Equation \ref{eq:seriesFit} to test 100 potential wind speeds between $0 \frac{m}{s}$ and $20 \frac{m}{s}$, and 100 potential wind directions between 0 degrees and 360 degrees. These parameters can be adjusted based on the user's need. For each tested wind speed and wind direction pair, a cost function determines how close of a match the guessed values are. The best matching wind speed and wind direction pair are the outputs of the algorithm. The wind direction is given in relation to the x-axis of the PSA body frame. To test this method, the data collected during the wind tunnel tests were input into the algorithm. Figure \ref{fig:algorithm8} shows the algorithm's outputs compared to the actual wind speed and wind direction in the wind tunnel. The largest wind direction error is 8.2 degrees, and the largest wind speed error is 2.7 $\frac{m}{s}$.

\section{Effects of 6-DOF Motion}
If the PSA is in motion, then the output of the algorithm is airspeed in the x-y plane. Translational motion in the x-y plane of the PSA can be subtracted from the airspeed to give wind velocity. To do this, inputs from a GPS module need to be available to determine translational velocity. Once the algorithm output is broken out into x and y components, the translational velocity components can be subtracted. Then the heading of the PSA is used to relate the wind velocity in the body frame to the inertial reference frame. The effects of velocity in the z-axis cannot be accounted for because Equation \ref{eq:seriesFit} only takes into account wind in the x-y plane.
\newline
\newline
\indent
 Although this has not been validated in the wind tunnel, Equation \ref{eq:seriesFit} could be used to remove the effects of roll pitch and yaw of the PSA. To do this, the inputs from an IMU module must be available to determine roll, pitch, and yaw of the PSA. The effects of yaw do not need to be removed as this is already accounted for in Equation \ref{eq:seriesFit}. Due to the symmetry of a typical pitot-static tube, the effects of the pitot-static tube's pitch can be described in the same way as the effects of the pitot-static tube's yaw. Also due to the symmetry, the roll of a pitot-static tube has no effect on its measurement. An educational paper performed a similar wind tunnel experiment but with only one pitot-static tube and the pitch was rotated 360 degrees instead of yaw.\cite{education} The normalized measured wind speed for each pitch angle match the values in Figure \ref{fig:normPlot}. This validates that the symmetry of a typical pitot-static tube allows the effects of both yaw and pitch to be described by Equation \ref{eq:seriesFit}. When the PSA pitches, the x-axis experiences pitch and the y-axis experiences roll. Likewise, when the PSA rolls, the y-axis experiences pitch and the x-axis experiences roll. If the effects of roll, pitch, and yaw are to be removed, they should be removed before the effects of translational motion are removed.

\section{Conclusion}
Proof of concept has been shown for the PSA sensor in combination with the algorithm presented as a method to estimate horizontal wind velocity in all directions. If used for a UAV, a specific flight path would not be required. The PSA sensor provides a light weight and cost effective alternative to existing devices that measure horizontal wind velocity. The exact limitations of a sensor of this kind still need to be further verified, and full scale UAV flight tests need to be performed. Further design iterations of the PSA can also be done to decrease it's payload and cost even more. 

\bibliographystyle{unsrt}
\bibliography{references}

\begin{thebibliography}{14}
\newcommand{\enquote}[1]{``#1''}
\providecommand{\natexlab}[1]{#1}
\providecommand{\url}[1]{\texttt{#1}}
\providecommand{\urlprefix}{URL }
\expandafter\ifx\csname urlstyle\endcsname\relax
  \providecommand{\doi}[1]{doi:\discretionary{}{}{}#1}\else
  \providecommand{\doi}{doi:\discretionary{}{}{}\begingroup
  \urlstyle{rm}\Url}\fi

\bibitem[1]{pisasale2003examining}
Pisasale, A., and Ahmed, N., \enquote{Examining the effect of flow reversal on
  seven-hole probe measurements,} \emph{AIAA journal}, Vol.~41, No.~12, 2003,
  pp. 2460--2467.

\bibitem[2]{pisasale2002novel}
Pisasale, A., and Ahmed, N., \enquote{A novel method for extending the
  calibration range of five-hole probe for highly three-dimensional flows,}
  \emph{Flow Measurement and Instrumentation}, Vol.~13, No. 1-2, 2002, pp.
  23--30.

\bibitem[3]{elston2015overview}
Elston, J., Argrow, B., Stachura, M., Weibel, D., Lawrence, D., and Pope, D.,
  \enquote{Overview of small fixed-wing unmanned aircraft for meteorological
  sampling,} \emph{Journal of Atmospheric and Oceanic Technology}, Vol.~32,
  No.~1, 2015, pp. 97--115.

\bibitem[4]{gracey1958summary}
Gracey, W., \enquote{Summary of methods of measuring angle of attack on
  aircraft,} 1958.

\bibitem[5]{ramakrishnan2007development}
Ramakrishnan, V., and Rediniotis, O.~K., \enquote{Development of a 12-hole
  omnidirectional flow-velocity measurement probe,} \emph{AIAA journal},
  Vol.~45, No.~6, 2007, pp. 1430--1432.

\bibitem[6]{matayoshi2005development}
Matayoshi, N., Inokuchi, H., Yazawa, K., and Okuno, Y., \enquote{Development of
  airborne ultrasonic velocimeter and its application to helicopters,}
  \emph{AIAA Atmospheric Flight Mechanics Conference and Exhibit}, 2005, p.
  6118.

\bibitem[7]{azartash2017evaluation}
Azartash-Namin, S.~K., \enquote{Evaluation of Low-Cost Multi-Hole Probes for
  Atmospheric Boundary Layer Investigation,} Ph.D. thesis, 2017.

\bibitem[8]{johansen2015estimation}
Johansen, T.~A., Cristofaro, A., S{\o}rensen, K., Hansen, J.~M., and Fossen,
  T.~I., \enquote{On estimation of wind velocity, angle-of-attack and sideslip
  angle of small UAVs using standard sensors,} \emph{Unmanned Aircraft Systems
  (ICUAS), 2015 International Conference on}, IEEE, 2015, pp. 510--519.

\bibitem[9]{mayer2012no}
Mayer, S., Hattenberger, G., Brisset, P., Jonassen, M.~O., and Reuder, J.,
  \enquote{A ‘no-flow-sensor’wind estimation algorithm for unmanned aerial
  systems,} \emph{International Journal of Micro Air Vehicles}, Vol.~4, No.~1,
  2012, pp. 15--29.

\bibitem[10]{langelaan2011wind}
Langelaan, J.~W., Alley, N., and Neidhoefer, J., \enquote{Wind field estimation
  for small unmanned aerial vehicles,} \emph{Journal of Guidance, Control, and
  Dynamics}, Vol.~34, No.~4, 2011, pp. 1016--1030.

\bibitem[11]{lisa}
Schibelius, L.~M., and Montalvo, C.~J., \enquote{Multi-MASS: A Fleet of
  Unmanned Aerial Vehicles for Atmospheric Characterization,} \emph{9th AIAA
  Atmospheric and Space Environments Conference}, 2017, p. 4475.

\bibitem[12]{sensirion}
Sensirion, \enquote{Directional Wind Meter Using SDP3x,} , 2017.
\newblock
  \urlprefix\url{https://developer.sensirion.com/applications/directional-wind-meter-using-sdp3x/}.

\bibitem[13]{windspeed}
Gracey, W., \enquote{Measurement of aircraft speed and altitude,} Tech. rep.,
  NATIONAL AERONAUTICS AND SPACE ADMINISTRATION HAMPTON VA LANGLEY RESEARCH
  CENTER, 1980.

\bibitem[14]{education}
Beck, B.~T., \enquote{AC 2010-1803: THE AERODYNAMICS OF THE PITOT-STATIC TUBE
  AND ITS CURRENT ROLE IN NON-IDEAL ENGINEERING APPLICATIONS,} \emph{age},
  Vol.~15, 2010, p.~1.

\end{thebibliography}

\end{document}